# Thermal Conduction of One-dimensional Carbon Nanomaterials and Nanoarchitectures*

Haifei Zhan( 占海飞)[1,2], and Yuantong Gu(顾元通)[2†]

[1]*School of Computing, Engineering and Mathematics, Western Sydney University, Locked Bag 1797, Penrith NSW 2751, Australia*
[2]*School of Chemistry, Physics and Mechanical Engineering, Queensland University of Technology (QUT), Brisbane QLD 4001, Australia*

This review summarizes the current studies of the thermal transport properties of one-dimensional (1D) carbon nanomaterials and nanoarchitectures. Considering different hybridization states of carbon, emphases are laid on a variety of 1D carbon nanomaterials, such as diamond nanothreads, penta-graphene nanotubes, supernanotubes, and carbyne. Based on experimental measurements and simulation/calculation results, we discussed the dependence of the thermal conductivity of these 1D carbon nanomaterials on a wide range of factors, including the size effect, temperature influence, strain effect, and others. This review provides an overall understanding of the thermal transport properties of 1D carbon nanomaterials and nanoarchitectures, which paves the way for effective thermal management at nanoscale.

## 1. Introduction

The advancement of nanotechnology has greatly driven the continuing miniaturization of electronic devices. The significantly increased levels of power density in the next generation electronic, optoelectronic, and photonic devices[1] demand an effective thermal management at nanoscale. Depending on the applications, the materials are required to have either a high thermal conductivity or a strongly suppressed thermal conductivity.[2, 3] In this regard, past decade has witnessed a great research efforts devoted to explore the thermal transport properties of nanomaterials, ranging from one-dimensional (1D) to three-dimensional (3D).[4]

As the building blocks for various engineering applications, low-dimensional (i.e., 1D and 2D) nanomaterials have received the most extensive attention. Researchers have studied the thermal transport properties of nanowires,[5, 6] single-walled or multi-walled carbon nanotube,[7, 8] BN, SiC, and Ge nanotubes.[9-13] It is found that the carbon nanotube (CNT) has a thermal conductivity as high as ∼ 3,000-3,500 W/mk at room temperature.[14, 15] Similarly, the thermal transport properties of a wide range of 2D nanomaterials have been studied,[16] such as graphene,[17-22] boron nitride,[23] silicene,[24, 25] phosphorene,[26] monolayer MoS$_2$,[27-29] and superlattice.[30] For instance, the phosphorene is found to possess a giant phononic anisotropy, which is squarely opposite to its electronic counterpart and can be effectively tuned by strain.[31] A large volume of works have also emphasized on the effective ways for thermal conductivity engineering, such as strain engineering,[32] structural defect engineering,[33] isotopes and functionalizations,[34-37] and others.[38] Besides, tremendous attempts have also been carried out to overcome the low thermal conductivity characteristic of 3D materials by taking low-dimensional nanomaterials as fillers. For example, adding CNT or graphene to nanocomposite, its thermal conductivity is effectively improved.[39-41]

---

* Project supported by Australian Research Council (ARC) Discovery Project DP170102861.
† Corresponding author. E-mail: yuantong.gu@qut.edu.au

Given the large variety of nanomaterials, carbon-based nanomaterials occupy a unique place in terms of the heat conduction as their thermal conductivity span an extraordinary large range - over five orders of magnitude.[2] In this paper, we aim to review the recent studies on the thermal transport properties of one-dimensional (1D) carbon nanomaterials and nanoarchitectures. We will first brief the big family of 1D carbon nanomaterials. Thereafter, emphasis will be laid on their thermal transport properties as acquired from simulations and experiments. In the end, some conclusions and perspectives will be given.

## 2. One-dimensional carbon nanomaterials and nanoarchitectures

Considering the hybridization states of carbon (sp$^3$, sp$^2$ and sp$^1$), there are many different 1D carbon nanomaterials and nanoarchitectures have been experimentally synthesized or theoretically predicted. These include the sp$^3$ carbon structures such as diamond nanowire and nanothread, the mostly studied sp$^2$ carbon structures – CNT, and the mixed sp$^1$ carbon structures like carbyne. Following will concisely introduce the characteristics of these 1D carbon nanostructures.

### 2.1. sp$^3$ bonded 1D carbon nanomaterial

The well-known sp$^3$ bonded 1D carbon structure is diamond nanowire or nanorod, which has been reported with unique features, e.g., good biocompatibility, negative electron affinity, and chemical inertness.[42] MD simulations show that diamond nanowire possesses excellent mechanical properties, e.g., Young's modulus around 688 GPa and yield strength about 63 GPa for <100> nanowire at 300 K.[43] Although tremendous efforts have been devoted, synthesis of diamond nanowires either through the reactive-ion etching (RIE) technique[44, 45] or chemical vapor deposition CVD method,[46] is still a challenge and ineffective, which greatly impedes their engineering applications.

Recently, researchers reported a new kind of ultra-thin sp$^3$ carbon structure – diamond nanothread (DNT), which is obtained from high-pressure solid-state reaction of benzene.[47] The DNT structure has two distinct sections, including the straight tube section as constructed from poly-benzene rings and the section with the so-called Stone-Walls (SW) transformation defects, as illustrated in Fig. 1.[48] The existence of SW transformation defects interrupts the central hollow of the structure, and differ them from the hydrogenated (3,0) CNTs. Considering this geometrical feature, we refer a DNT unit cell with $n$ poly-benzene rings between two adjacent SW transformation defects as DNT-*n* for discussion convenience, e.g., DNT-8 has eight poly-benzene rings with a length approximating 4 nm (Fig. 1). Following the experimental success, density functional theory (DFT) calculations show that there are 15 different stable structures can be constructed by considering all possible bonding geometries within a one-dimensional stack of six-fold rings.[49, 50] These stable DNTs can be divided into three groups based on their structural properties, including achiral, stiff chiral and soft chiral, which also cover the previously proposed $sp^3$ nanothreads, i.e., tube (3,0),[48] polymer I,[50] and polytwistane.[51, 52]

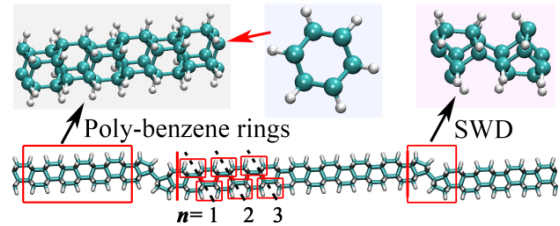

**Fig. 1.** (color online) Schematic view of the DNT as synthesised from experiments. (a) A segment of the DNT. Top panel shows the structural representation of the poly-benzene rings (left and middle insets) and the Stone-Wales transformation defects (SWD, right inset). Adapted from Ref. [53] –Published by The Royal Society of Chemistry.

Despite certain deviations, both DFT calculations and MD simulations have shown that DNT has excellent mechanical properties.[54] For the theoretically predicted nanothreads, DFT calculations show that they have a wide range of Young's modulus ranging from 0.08 to 1.16 TPa.[49] Based on MD simulations (using reactive force field - ReaxFF[55]), researchers found that the experimentally synthesised DNT has a high tensile stiffness comparable with that of CNT (i.e., around 850 GPa), and its yield strain is approximately 14.9% with an ultimate stress of about 143.3 GPa. Following works (using adaptive intermolecular reactive bond order - AIREBO potential[56]) show that the yield strength of DNT is almost independent of the constituent unit cells, and its failure behavior is controlled by the SW transformation defect section.[57] More interestingly, the ductility of DNT is found tunable by altering the number of SW transformation defects. That is the DNT with less SW transformation defects has smaller yield strain and is more brittle. MD simulations have also been extended to probe the mechanical properties of the theoretically predicted 15 different DNTs, and their mechanical properties are observed heavily dependent on their morphology, temperature,[58] and hydrogenation states.[59] Further studies show that the excellent mechanical properties of DNTs are well retained by introducing various surface functional groups.[60] Very recent works show that DNT can be used as effective reinforcements for nanocomposites,[61] and has strong interfacial load transfer capability for nanofiber applications.[62]

## 2.2. $sp^2$ bonded 1D carbon nanoarchitecture

Since its first discovery,[63] CNTs have attracted extensive research focuses, and they have been reported with tremendous applications owing to their excellent physical and chemical properties. Due to the large volume of review works existing in literature,[64, [65] in this work, we will focus on the CNT-based architectures.

According to the geometrical characteristics, we can classify the $sp^2$ bonded 1D carbon nanomaterials into two groups, including the CNT analogous and CNT derivatives. Geometrically, CNT can be regarded as a tube structure wrapped up from a graphene sheet. Therefore, based on various derivatives of graphene, different kinds of $sp^2$ bonded tube structures have been theoretically proposed. Fig. 2a illustrates two representative CNT analogous, including the nanotube as constructed from penta-graphene,[66] and the nanoscroll.[67] As the name indicates, penta-graphene is composed entirely of carbon pentagons, which can be exploited from T-12 carbon.[66] MD simulations (using AIREBO potential[56]) show that the penta-graphene nanotube possesses a very high ductility (with failure strain exceeding 60%) due to the irreversible pentagon-to-polygon structural transformation.[68] EMD simulations predict that penta-graphene has a thermal conductivity around 167 W/mK.[69] The first successful synthetization of nanoscroll was reported in 2003,[70] which has been reported with promising applications for $H_2$ storage. Following MD simulations show that by the aid of out-of-plane force as induced by the hydrogenation or fluorination, the graphene morphology can be well manipulated to form nanoscroll.[71, 72] It is found that the nanoscroll crystal exhibits strong hysteresis under compression, which shows promising usage as energy-absorbing material.[73] Besides the tube structure, it is worthy to mention another kind of 1D nanostructure – helical nanostructure.[74] Inspired by biological materials, the formation and mechanical properties of the helical nanostructure has been extensively studied,[75-77] though their thermal transport properties have been rarely touched.

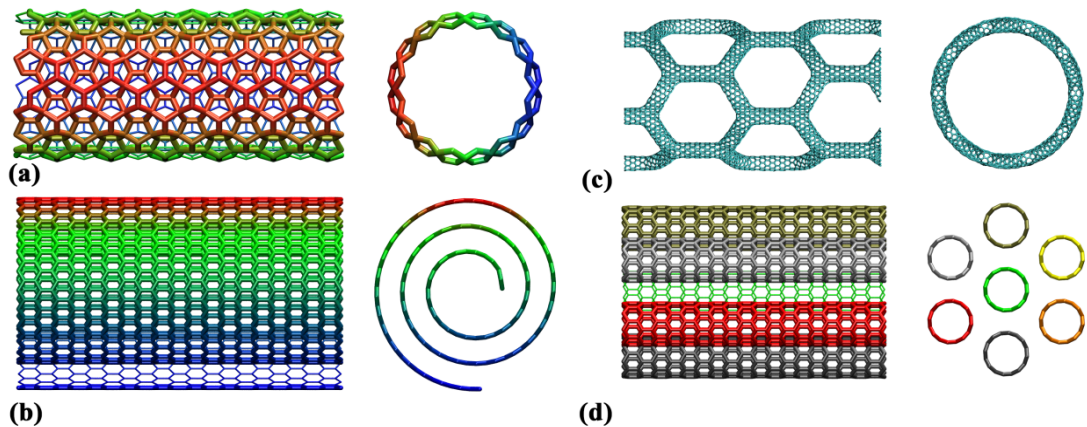

**Fig. 2.** (color online) $sp^2$ bonded carbon nanostructures. Atomic configurations of CNT analogous: (a) penta-graphene nanotube, and (b) nanoscroll; Atom configurations of CNT derivatives: (c) supernanotube, and (d) nanofiber. In each figure, left panel is the front view, and the right panel is the cross-sectional view.

Fig. 2b shows another category of $sp^2$ bonded nanoarchitecture, including supernanotube and nanofiber, which we refer as CNT derivatives as these structures take CNT as the building blocks. Experimental success in forming various single-walled CNT junctions (e.g., T-, Y-, and X-shaped) have greatly promoted the construction of CNT-based nanoarchitectures,[78, 79] e.g., supergraphene,[80] CNT-pillared graphene,[81, 82] and supernanotube. Plenty of theoretical works have studied the stability and formation mechanisms of CNT junctions, and the design of supernanotubes based on these junctions.[83, 84] Specifically, the supernanotube is reported to exhibit metallic and semiconducting behaviour,[84] and high flexibility.[85] MD simulations show that Young's modulus of the supernanotube is inversely proportional with its radius and dependent on its charity. Also, it deforms like a fishing net under tensile load.[86] Carbon nanofiber is another CNT-based nanoarchitecture being widely studied, which can be fabricated through either spinning[87, 88] or twisting/rolling techniques.[89] CNT fibers have been reported with outstanding mechanical, chemical and physical properties and well over-perform traditional carbon and polymetric fibres.[90-95] Tremendous appealing applications have been proposed for CNT fibers, including next-generation power transmission lines, artificial muscles,[96, 97] aerospace electronics and field emission,[90] batteries,[98] intelligent textiles and structural composites.[99]

## 2.3. sp$^1$ bonded 1D carbon architecture

In literature, there are two kinds of nanostructures containing sp$^1$ bonded carbons, including the so-called graphyne and graphdiyne (2D),[100, 101] and the 1D monoatomic chain – carbyne. Although the synthesis of graphyne has not been realized, researchers have already reported the successful synthetization of graphdyine.[102] Several studies have reported the mechanical,[103, 104] thermal properties[105, 106] and electronic properties of graphyne or graphdiyne.[107] Apparently, by rolling up a graphyne or graphdiyne sheet, a tube can be formed similar as that of CNT (see Fig. 3a). A recent work has reported the successful synthetization of graphdiyne nanotube array, though the thickness is large - around 15 nm.[108]

In the other hand, the 1D carbon chain has gained substantial interests due to its promising usage in energy storage devices and nanoscale electronics. Basically, carbyne is a 1D sp-hybridized carbon allotrope which has two forms including α-carbyne (polyyne, with alternating single and triple bonds) and β-carbyne (cumulene, with repeating double bonds), as schematically shown in Fig. 3b. Though the existence of this ultrathin carbon has been debated,[109] several experimental works have reported its synthetization or fabrication.[110-112] Recent work shows that cumulene can transition to polyyne at the temperature of 499 K, and the transition is very fast within 150 fs.[113] Both MD simulations[114] and first principle calculations[115] show that carbyne has excellent mechanical properties, e.g., a high stiffness that requires a force of ~ 10 nN to break. According to the theoretical calculations based on the nonequilibrium Green's function method and density functional theory, carbyne is found to have wire length-dependent conductance.[1]

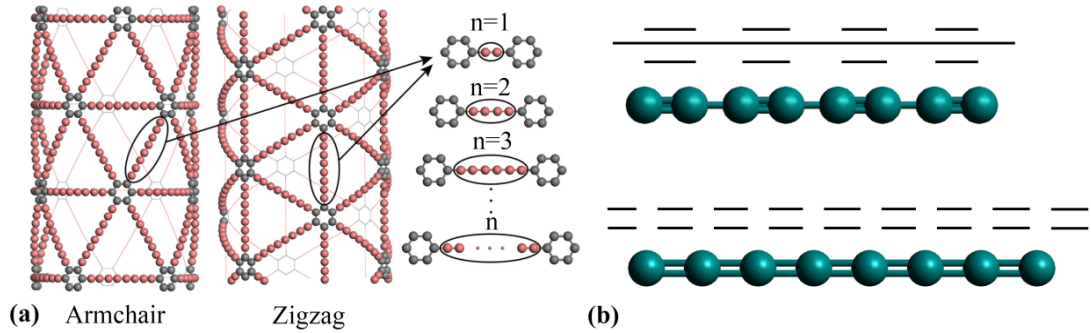

**Fig. 3.** (color online) sp[1] bonded 1D carbon nanostructures. (a) Atomic configurations of the graphdiyne nanotube; Adapted with permission from Ref. [116]. Copyrighted by the American Physical Society. (b) Atomic configurations of the monoatomic chain – polyyne (top panel) and cumulene (bottom panel).

## 3. Thermal transport properties

Above sections have introduced the main 1D carbon nanoarchitectures and nanomaterials, following section will focus on the thermal transport properties of those materials. The thermal conductivity of carbon nanomaterials can be measured through experiments (using the steady state or transient 3ω techniques),[2] or estimated from simulations or first principle calculations. Molecular dynamics (MD) simulation is one of the most frequently applied simulation approaches, which includes equilibrium MD (EMD) simulation, non-equilibrium MD (NEMD) simulation, and reverse NEMD simulation or the Müller-Plathe method.[117] A comprehensive comparison of these simulation techniques can be found from elsewhere.[118, 119]

It is necessary to point out that the thermal conductivity as obtained from MD simulations relies heavily on the atomic potentials being applied. For the carbon system, there are many types of empirical potentials available, such as the most widely used AIREBO potential,[56] Tersoff potential,[120] ReaxFF[55] and charge-optimized many body (COMB) potential.[121] Since the empirical potential development is normally based on a set of database, and there are always certain trade-offs during the fitting process. As such, different empirical potentials have their own accuracy in describing the interested properties of the specific carbon systems, which results in unavoidable deviations for the calculated thermal conductivity in literature. A detailed comparison of the merits or demerits of the currently available carbon empirical potentials can be from Ref. [122].

### 3.1. sp[3] bonded 1D carbon nanomaterial

As aforementioned, the chemical inertness of sp[3]-hybridized carbon makes the synthesis of diamond nanowire or nanothread still a challenge. To date, most of the studies on the thermal transport properties of diamond nanowire or nanothread are carried out by MD simulations or first principle calculations.

#### 3.1.1. Thermal transport properties of diamond nanowire

The increased phonon-boundary scattering[123] or changes in the phonon dispersion[124]

leads to a suppressed heat conduct capability for nanowire compared with bulk crystals. Both MD simulation and *ab initio* calculations show that the diamond nanowire has a strong crystallographic orientation dependent thermal conductivity ($\kappa$), whereas *ab initio* calculations show a strongest heat conduction along <001> direction,[125] but NEMD simulations (using Tersoff and Brenner empirical potential[126]) predict the <011> crystal direction.[127, 128]

Recent MD simulations show that thermal conductivity of diamond nanowire (with surface functionalization) is around a factor of 4 smaller than that of the pristine (10,10) CNT,[129] which experiences much less influences from surface functionalization compared with that of the CNT. More importantly, $\kappa$ exhibits a strong dependence on the length and radius of the nanowire.[128, 129] For instance, $\kappa$ is found to increase from ~ 10 to ~ 25 W/mK when the length of the diamond nanowire increases from 3 to 8 nm. More interestingly, it is found that $\kappa$ of thicker diamond nanowire decreases with increasing temperature due to phonon-phonon scattering, while thinner nanowire appears insensitive due to the increased role of surface scattering.[128]

### 3.1.2. Thermal transport properties of DNT

Currently, studies of the thermal transport properties of DNT are still focused on the experimentally reported structure (see Fig. 1). Based on the NEMD simulation (using AIREBO potential[56]), the SW transformation defects (that interrupt the continuity of the tube structure) is found to induce the "interfacial thermal resistance" or Kapitza resistance (KR).[57] As illustrated in Fig. 4a, a clear temperature jump (around 4.2 K) is detected at the region with SW transformation defect (between the two tube sections). Comparing with CNT, the DNT (with one SW transformation defect) is estimated with a low thermal conductivity about 35.6 ± 4.7 W/mK (which is in the same order as that of the diamond nanowire). Here, the temperature gradient along the heat flux direction is approximated as $\Delta T/L$, with $\Delta T$ and $L$ representing the temperature difference between the heat source and sink, and length of the DNT, respectively. Calculations show that the DNT has significant reduction in phonon lifetimes almost over the entire range of frequency compared with that of the pristine (3,0) CNT (see Fig. 4b). Such reduction is supposed as originated from the additional phonon scattering at the SW transformation defect, which eventually suppress the thermal conductivity of DNT.

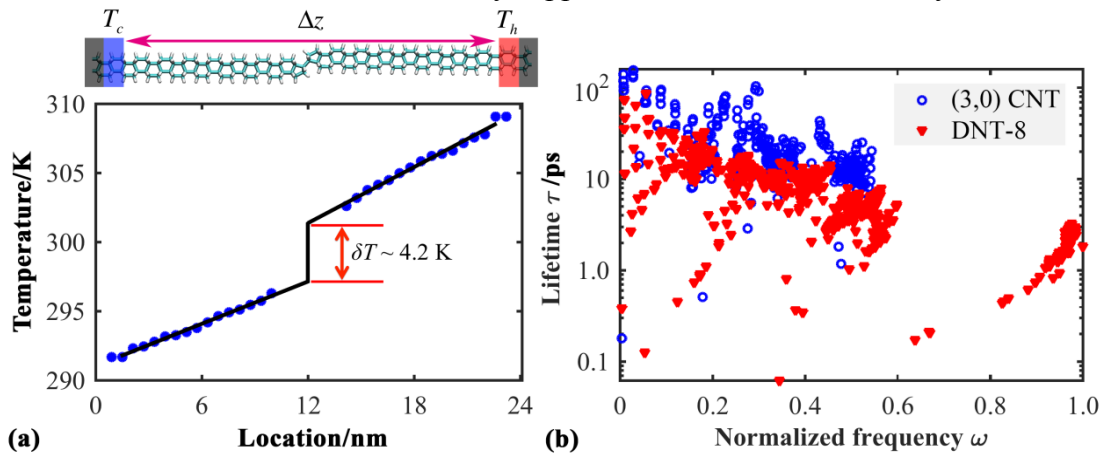

**Fig. 4.** (a) Temperature profile of DNT-55 (length of ∼ 24 nm), which possesses only a single SW transformation defect at the middle of the structure. Top panel schematically show the studied sample. (b) Comparisons of the mode lifetimes between (3,0) CNT and DNT-8. The mode lifetime is calculated using the package Jazz[130, 131] based on the corresponding vibrational density of states (VDOS). Adapted from Ref. [57], with permission from Elsevier.

The most intriguing observation is that the DNT exhibits a superlattice thermal transport characteristic.[57] Similar as observed from diamond nanowire or silicon nanowires, $\kappa$ of DNT increases when the sample length increases, which is uniformly observed for the DNT with different number or density of SW transformation defects (see Fig. 5a). Particularly, $\kappa$ of the DNT with a given sample length initially decreases for smaller number of poly-benzene rings, and then undergoes a relatively smooth increase. By normalizing $\kappa$ with its minimum value, a consistent scaling behaviour is observed (Fig. 5b). Similar feature are also reported for thin-film,[132] nanowire[133] and superlattices,[134, 135] which can be explained from the perspective of phonon coherence length. Specifically, the phonon coherence length of DNT can be calculated from $L_{co}=\tau_{co}v_D$. Here, the coherence time $\tau_{co}$ can be estimated from the normalized VDOS according to the Bose-Einstein distribution $D_{BE}$, and the Debye velocity $v_D$ is derived from the sound velocity for the one longitudinal and two transverse branches, respectively.[136, 137] Estimation shows that the DNT possess a coherence length around 5.6 nm, corresponding to the unit length of DNT-11 and agrees well with the result in Fig. 5b. It is concluded that for DNT with poly-benzene number smaller than around 10, phonon transport is largely dominated by wave effects including constructive and destructive interferences arising from interfacial modulation. In comparison, when more poly-benzene rings (or less SW transformation defects) are presented, phonon waves lose their coherence and transport is more particle-like.

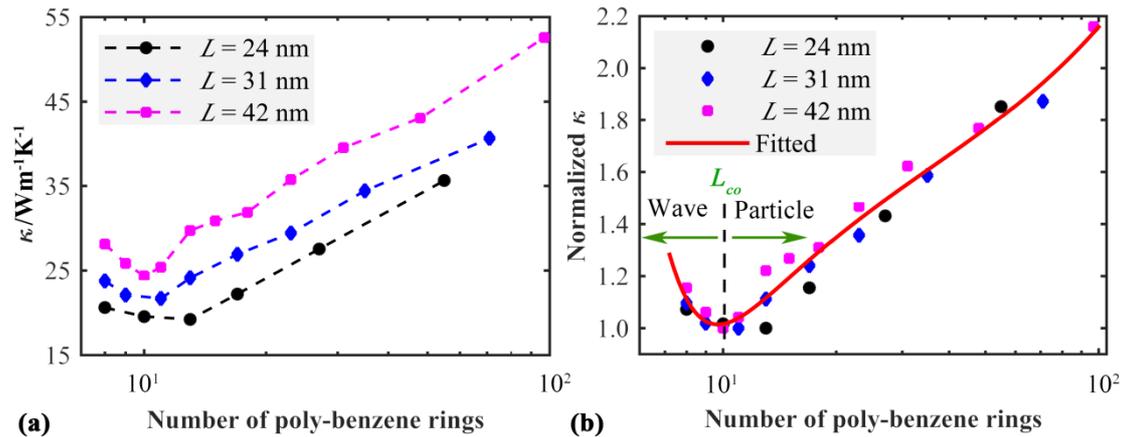

**Fig. 5.** (a) Thermal conductivity of DNT as a function of length and the constituent unit cells (with different number of poly-benzene rings); (b) Normalized thermal conductivity (by its respective minima) shows a general valley trend. Reprinted from Ref. [57], with permission from Elsevier.

Similar as observed from other 1D nanostructures,[5, 80, 138, 139] 2D nanoribbons,[29, 140] and bulk materials,[118] $\kappa$ of DNT also exhibits a strong length dependency. As shown in Fig. 6a, the inverse of $\kappa$ follows a linear scaling relationship with the inverse of its length, which is in line with the linear relationship as derived from the kinetic theory,[118, 138] i.e., $1/\kappa = 1/\kappa_{\infty}(1 + \lambda/L)$. Here, $\kappa$ and $\kappa_{\infty}$ are the size dependent and converged (when sample size is large enough) thermal conductivity, respectively. $L$ is the sample length, and $\lambda$ is the mean-free-path (MFP). For the DNT-8, the $\kappa_{\infty}$ and MFP are estimated as 55.18 W/mK and 40.18 nm, respectively. Additionally, $\kappa$ of DNT gradually decreases when the temperature increases. For the examined temperature ranging from 50 to 350 K, $\kappa$ roughly follows a power-law relationship with the temperature. According to the Boltzmann transport equation, the phonon thermal conductivity is a function of the specific heat, phonon group velocity and phonon relaxation time.[141] The phonon relaxation time is mainly limited by boundary scattering and the three-phonon umklapp scattering processes.[142] At high temperature, the role of umklapp scattering is dominating, and the scattering rate of the umklapp process is proportional to temperature.[143, 144] For homogeneous materials, $\kappa$ is expected to drop steadily with the increasing temperature and obeys a power-law dependence with an exponent of -1. Due to the additional phonon scattering at the region with SW transformation defect, $\kappa$ of DNT exhibits a weaker power-law relationship with temperature. Further EMD simulations show that $\kappa$ of defected DNT experiences a factor of ~ 5 reduction across the temperature from 200 – 1000 K.[145]

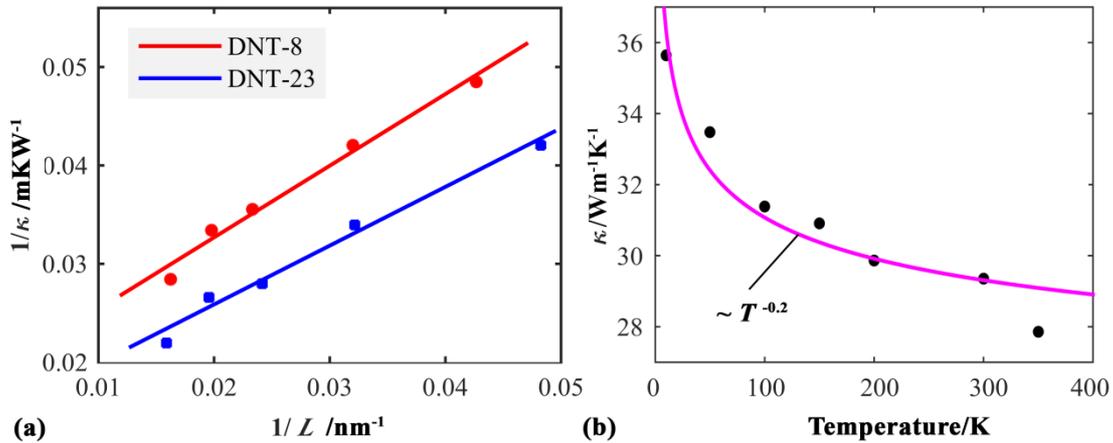

**Fig. 6.** (a) The linear scaling relationship between the inverse of $\kappa$ and the inverse of sample length. (b) Thermal conductivity of DNT-13 (size of ~ 42 nm) as a function of temperature. The solid line is fitted with a power-law relation. Adapted from Ref. [57], with permission from Elsevier.

## 3.2. sp² bonded 1D carbon nanoarchitecture

Since there are plenty of works already summarized the thermal transport properties of CNT, which thus will not be repeated herein. It is found that although several works have investigated the thermal conductivity of penta-graphene,[69, 146, 147] the penta-graphene nanotube is still untouched. Similar as the carbon nanoscrolls, few studies have reported their thermal transport properties.[148] As such, in this section, we will review the recent works on the thermal transport properties of CNT derivatives including CNT fibers and supernanotubes.

### 3.2.1.  Thermal transport properties of CNT fibers

Thermal transport property of CNT fibers is one of the properties being extensively studied due to their promising applications in nanoscale electronics, such as supercapacitor electrodes.[149] Although the experimental measurements are not consistent, the CNT fibers are showing good thermal transport properties in the axial direction as inherited from CNT.[150] For instance, a room-temperature thermal conductivity of $1,750 - 5,800$ W/mK is measured for the single-walled CNT rope.[151] By varying the density from $0.81$ g/cm$^3$ to $1.39$ g/cm$^3$, $\kappa$ of CNT arrays are found to increase from 472 to 766 W/mK.[152] Similar as other nanomaterials, the axial $\kappa$ exhibits a strong temperature dependence, i.e., it decreases smoothly with decreasing temperature, and displays linear temperature dependence below 30 K.[151]

Besides the axial thermal transfer, the inter-tube thermal conductivity of CNT fibers has also gained substantial attention. Due to the weak van der Waals (vdW) inter-tube interaction, the thermal conductivity is extremely low (e.g., around 0.35 W/mK from NEMD simulation[153]). Both experimental works and MD simulations have been carried out to explore the effective ways to enhance the inter-tube thermal conductivity, e.g., increase the density,[154] use ion irradiation.[155] According to the NEMD simulation (using AIREBO potential[56]), the inter-tube $\kappa$ can be effectively enhanced by compressive strain (before the buckling of CNT).[153]

### 3.2.2.  Thermal transport properties of supernanotubes

As discussed earlier, there are different kinds of CNT junctions can be utilized to build supernanotubes. Similar to single-walled CNT, the supernanotube (ST) can also be denoted by two geometrical index $N$ and $M$. Based on reverse NEMD simulations, the ST is found to exhibit a low $\kappa$, e.g., $\sim 5.4$ W/mK for ST(5,0) at 300 K, which varies with the junction type.[139] As illustrated in Fig. 7a, the inverse of $\kappa$ follows a linear relationship with the inverse of the sample length, same as commonly observed from other nanomaterials. By extrapolating the linear relationship to infinite sample length, the ST(5,0) is supposed to have a saturated $\kappa$ of $\sim 5.9$ W/mK. It is found that $\kappa$ decreases when the ST's diameter increases, which is in line with that of single-walled CNT.[142] Such observation is due to the fact that the increased diameter will decrease the average group velocity and augment the probability of umklapp process, which thus leads to a suppressed $\kappa$. Similar as CNT,[15, 141] the thermal conductivity of ST shows a proportional relationship with the inverse of temperature. According to Fig. 7b, $\kappa$ decreases continuously when the temperature increases from 200 to 500 K.

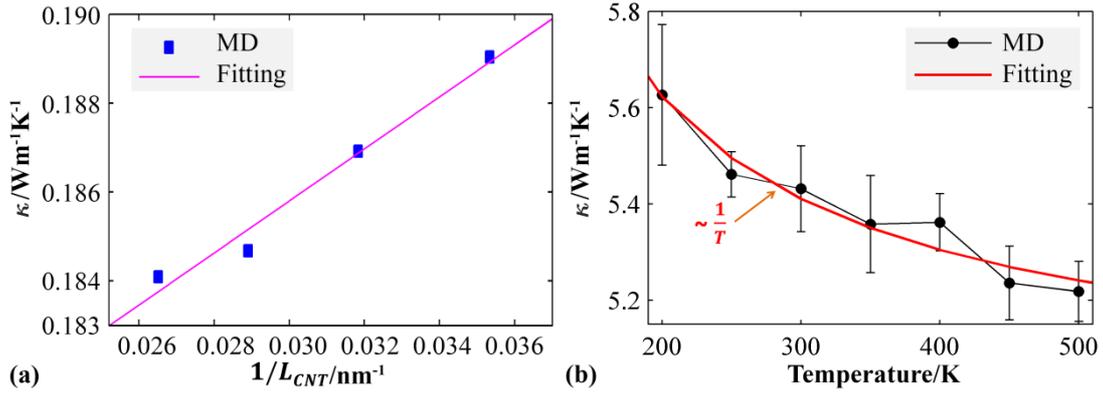

**Fig. 7.** Thermal conductivity of supernanotube. (a) Thermal conductivity as a function of sample length; (b) Thermal conductivity of ST(5,0) as a function of temperature. The solid line is fitted with a power-law relation. The constituent unit is (8,0) single-walled CNT. Adapted from Ref. [139], reproduced by permission of the Royal Society of Chemistry.

Although $\kappa$ of ST exhibits a similar relationship with size and temperature as that of CNT, it has a totally different relationship with axial strain. As compared in Fig. 8a, the relative $\kappa$ of the ST(5,0) fluctuates around one, signifying the insignificant impacts from the axial strain on its thermal conductivity. In comparison, for a (8,0) SWNT, the relative $\kappa$ decreases continuously when the strain increases from negative (compression) to positive (tension). Such strain insensitive characteristic of the $\kappa$ of the ST can be explained from the fact that when the ST is under axial strain, its constituent CNTs will be either compressed or stretched depending on their orientations. Considering the tensile scenario, the constituent CNT along the loading direction is under tensile while the CNT along the circumferential direction is under compression (see inset of Fig. 8a). In other words, the enhancement and degrading effect from the compressed and stretched CNTs compensate to each other and thus makes the thermal conductivity of the ST insensitive to the axial strain. Further evidence is seen from their VDOS. As compared in Fig. 8b, the axial stain softens the high-frequency phonon modes of the CNT (i.e. G-band phonon modes), which decreases the specific heat and thus leads to a continuous reduction to $\kappa$. However, the low-frequency and high-frequency phonon modes are almost unchanged for the ST at different strains.

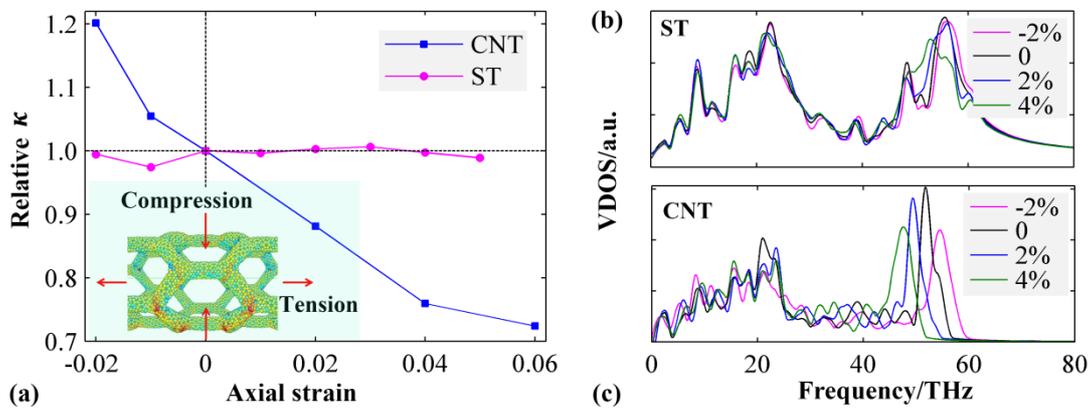

**Fig. 8.** (a) Relative $\kappa$ of the ST(5,0)@(8,0)-$\gamma$ and SWNT as a function of axial strain.

Comparisons of the VDOS at different axial strains for: (b) ST and (c) single-walled CNT. Reprinted from Ref. [139], reproduced by permission of the Royal Society of Chemistry.

### 3.3. sp¹ bonded 1D carbon nanomaterial

Similar as the $sp^3$ bonded 1D carbon nanomaterials, the synthetization of $sp^1$ bonded 1D carbon nanomaterials is also still a challenge. Thus, the current understanding of its thermal transport properties is mainly established by MD simulations or first principle calculations.

### 3.3.1. Thermal transport properties of graphyne nanotube

Based on NEMD simulations (using AIREBO potential[56]), $\kappa$ of graphyne nanotube exhibits an unprecedentedly low thermal conductivity (below 10 W/mK) at room temperature.[116] Same results are obtained from EMD simulations (using AIREBO potential[56]).[156] According to the phonon polarization and spectral energy density analysis, such ultra-low thermal conductivity originals from the large vibrational mismatch between the weak acetylenic linkage and the strong hexagonal ring. Specifically, the thermal conductivity is found to experience a sharp decrease when the number of acetylenic linkages increases (smaller than 5). For larger number of acetylenic linkages, such decrease trend is largely slowed. Additionally, the thermal conductivity of graphyne nanotube is found to exhibit a linear scaling relationship with its length. It is observed that the nanotube with more acetylenic linkages has a weaker length dependence (see Fig. 9a).[116, 157] Further studies show that $\kappa$ is independent of the diameter when the nanotube has 10 acetylenic linkages. For the counterpart with only one acetylenic linkage, $\kappa$ decreases with diameter (when the diameter is smaller than ~ 2 nm, see Fig. 9b).

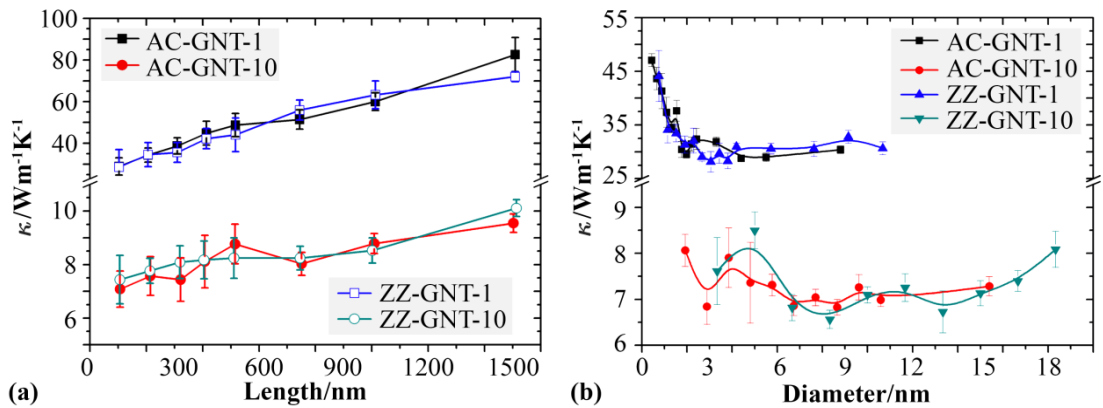

**Fig. 9.** (color online) Thermal conductivity of graphyne nanotube (GNT). Thermal conductivity as a function of: (a) sample length; and (b) sample diameter. AC and ZZ refer to armchair and zigzag, respectively. GNT-1 and GNT-10 refer to the nanotube with one and ten acetylenic linkages, respectively. Adapted with permission from Ref. [116]. Copyrighted by the American Physical Society.

### 3.3.2. Thermal transport properties of carbyne

Consistent with other nanomaterials, the thermal conductivity of carbyne as obtained from reverse NEMD simulations (using ReaxFF potential[55]) exhibits a strong length dependency. That is, $\kappa$ increases from ~ 200 to 680 W/mK when the sample length increases from 20 to 40 nm. Similar as observed from CNT, $\kappa$ decreases when the axial strain varies from compressive to tensile. Specifically, a small tensile strain (~ 3%) is found to induce ~ 70% decrease in the thermal conductivity. Other study based on the EMD simulations (using polymer consistent force filed - PCFF potential[158]) shows that $\kappa$ of carbyne shows a positive temperature dependence at low temperature range and becomes negative at higher temperature range (due to the increased phonon-phonon scattering when more phonons are excited at higher temperature).[159]

It is worthy to mention that cumulene and polyyne has totally different thermal transport properties. Based on EMD simulation (using ReaxFF potential[55]), the thermal conductivity for metallic cumulene (with 50 carbon atoms) is about 83 W/mK at 480 K.[113] Estimations show that the electronic contribution is around $1.11 \times 10^{-3}$ W/mK, which is ignorable compared with the phonon thermal conductivity. Unlike cumulene, the polyyne chain exhibits a much smaller thermal conductivity, around 42 W/mK at 500 K, which dramatically decreases to 5.5 W/mK when only 4% defective bonds are presented. Here, the defect in polyyne chain refers to the carbon bond (~ 1.45 Å) that is shorter than single bond but longer than double and triple bonds. Such defect is observed during the formation process when the heating rate is high than around 2.5 K/fs or if the cumulene chain is initialized at a temperature over 499 K in the DFT calculations.[113] The underlying mechanism of the remarkable thermal transport difference among cumulene, polyyne, and defected polyyne can be explained by comparing the heat current autocorrelation function (HCACF). Theoretically, the decay of HCACF in bulk material is exponential, and the high-frequency phonon modes are responsible for the initial fast decay while slow decay arises from low-frequency phonon modes.[113] As compared in Fig. 10a, the cumulene exhibits a rapid initial decay followed by a long time decay (around 15 ps), whereas, the polyyne has a shorter decay time (about 10 ps). Since high-frequency phonons have a limited contribution to the phonon thermal conductivity (due to their low group velocity), the remarkable difference in the thermal conductivity of cumulene and polyyne is attributed to the large difference in relaxation time for long-wavelength phonons along the chain. In comparison, the HCACF of defective polyyne is much lower than that of the pristine polyyne (see Fig. 10b), and the decay time is remarkably shorter. It is regarded that the existences of defective carbon bonds increase phonon scattering, which results in localized vibrational modes that suppress the thermal conductivity.

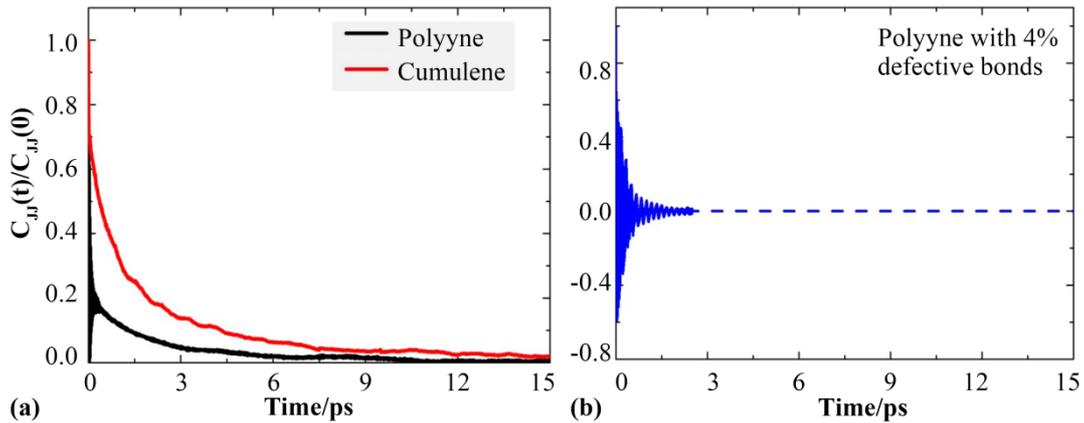

**Fig. 10.** (color online) Heat flux autocorrelation (HFAC) functions for carbyne chains. (a) Cumulene and perfect polyyne chains; (b) Polyyne chain with 4% defective bonds. Adapted with permission from Ref. [113]. Copyright 2017 American Chemical Society.

## 4. Conclusion and perspectives

Different hybridization states of carbon have provided us a wide opportunity to construct a variety of 1D nanomaterials and nanoarchitectures, such as diamond nanowires, diamond nanothreads, carbon nanotubes, nanoscrolls, supernanotubes, and carbyne. Extensive works have studied their thermal conductivity, and its dependence on various factors including the sample size (e.g., length or diameter), temperature, and mechanical strain. The strong dependence of the thermal conductivity on these factors offers an effective avenue to engineering the thermal transport properties of these 1D nanomaterials.

It is noted that majority of the current studies have focused on the $sp^2$ bonded 1D carbon nanostructure (though the helical nanostructures as inspired from biological materials have been rarely discussed). From the experimental perspective, the lacking of investigation on the mixed $sp^1$ or $sp^3$ bonded 1D carbon nanomaterials is originated form the huge challenges to synthesize them. In the other hand, MD simulations are seen to lead relatively large deviations in estimating the thermal conductivity due to the inaccuracy of the empirical potentials in describing mixed hybridized carbon bonds in $sp^1$ or $sp^3$ bonded 1D carbon nanomaterials.[116] Given the broad promises of these 1D carbon nanomaterials, a significant progress is expected for either the experimental synthetization or establish a more accurate empirical potential to describe them.

It is worthy to mention that there are various simulation approaches available for the investigation of nanoscale thermal phenomena, including Boltzmann transport equation (BTE), non-equilibrium Green's function (NEGF), and MD method. Each theoretical calculation/simulation method has its own merits and demerits. For instance, BTE method is a classical theory that applicable for diffusive regime, in which phonons are treated as quasi-particles. In contrast, NEGF method is normally adopted to probe the quantum or ballistic thermal conductance. For instance, NEGF method has been employed to study the quantum thermal conductance of CNT, [160, 161] while BTE method has been used to acquire the thermal conductivity of 2D materials in the diffusive limit.[162] Technically, it is challenging to use either BTE or NEGF method to study the thermal conduction of nanoarchitectures due to the large number of interfaces and changing dimensions within the system, e.g., junctions formed from 1D and 2D components. On the other hand, MD method does not rely on any thermodynamic-limit assumption, and it is thus applicable to model nanoscale systems with complex geometries in a straightforward way. Therefore, most of the theoretical works on thermal conductivity of nanoarchitectures are based on MD simulations due to the existing interfaces and structural complexities.

## Acknowledgment

This study is supported by the High Performance Computer resources at Queensland University of Technology.